\documentclass[12pt,a4]{article}
\pdfoutput=1
\usepackage{graphicx}
\usepackage{amsmath}
\usepackage{subfigure}
\usepackage{gensymb}
\usepackage[left=1.5in, right=1.5in, top=1.25in, bottom=1.25in]{geometry}
\date{10 March 2017}

\begin{document}

\title { \bf Search for the differences in  Atmospheric Neutrinos and Antineutrinos oscillation parameters at the INO-ICAL Experiment}

\author{Daljeet Kaur$^{\ast}$, Zubair Ahmad Dar$^{\ddagger}$, Sanjeev Kumar$^{\dagger}$, Md. Naimuddin$^{\dagger}$ \\
$^{\ast}$S.G.T.B. Khalsa college, University of Delhi\\
$^{\ddagger}$Aligarh Muslim University, Aligarh\\
$^{\dagger}$Department of Physics and Astrophysics, University of Delhi}

\maketitle

\begin{abstract}
In this paper, we present a study to measure the differences between the atmospheric neutrino and anti-neutrino oscillations in the Iron-Calorimeter detector at the India-based Neutrino 
Observatory experiment. Charged Current $\nu_{\mu}$ and $\overline{\nu}_{\mu}$ interactions with the detector under the influence of earth matter effect have been simulated for ten years of exposure. 
The observed $\nu_{\mu}$ and $\overline{\nu}_{\mu}$ events spectrum are separately binned into direction and energy bins, and a $\chi^{2}$ is minimised with respect to each bin to extract the oscillation 
parameters for $\nu_{\mu}$ and $\overline{\nu}_{\mu}$ separately. We then present the ICAL sensitivity to confirm a non-zero value of the difference in atmospheric mass squared of neutrino and anti-neutrino 
i.e. $|\Delta m^{2}_{32}|-|\Delta\overline{m^{2}}_{32}|$. 
\end{abstract}

\newpage

\newpage 

\section{Introduction}
 In the last few decades, neutrino oscillation experiments have provided many model independent evidences of neutrino oscillations. It all started in 1998 when Super-Kaimokande observed that the 
 atmospheric  muon neutrinos are changing flavor as they traverse through the atmosphere \cite{SK1, SK2}. In 2001, the Sudbury Neutrino Observatory experiment obtained a direct evidence of a flavor 
 change of solar neutrinos due to  neutrino oscillations \cite{SNO1}. Next, KamLand experiment observed the same effect with reactor neutrinos in 2002 \cite{KAM1}. There are now multiple next generation 
 of experiments aimed at studying neutrino oscillation  using different neutrino sources. The neutrino  oscillations within the three flavor framework is well established from solar, atmospheric and 
 reactor neutrino experiments, and the oscillation parameters  are getting measured with better precision. The evidence of non-zero masses of neutrinos establish the fact that the three flavors of 
 neutrinos are mixed. In the three-flavor oscillation paradigm, neutrino  mixing can be described by a 3$\times$3 unitary mixing matrix known as Pontecorvo-Maki-Nakagawa-Sakata (PMNS) 
 matrix \cite{bruno, pmns1}. The three flavor eigenstates of neutrinos are mixtures of three mass eigenstates  according to the PMNS matrix. Under the standard parameterization of PMNS matrix, 
 the neutrino oscillation probabilities are defined in terms of three mixing angles $\theta_{12}$, $\theta_{23}$, $\theta_{13}$; two mass-squared differences $\Delta m^{2}_{21}$, $\Delta m^{2}_{32}$ and 
 a Dirac CP-violation phase $\delta_{CP}$. The resulting oscillation probabilities depends on the  these oscillation parameters (mixing angles and mass squared differences). For a given neutrino  of 
 energy $E_{\nu}$ and the propagation length L, the survival probability for $\nu_{\mu}$ is
 given by
\begin{equation}
P(\nu_{\mu} \rightarrow \nu_{\mu}) \simeq 1-4\cos^{2}\theta_{13}\sin^{2}\theta_{23} \times [1-\cos^{2}\theta_{13}\sin^{2}\theta_{23}]\sin^{2}(\frac{1.267 |\Delta m^{2}_{32}|L}{E_{\nu}}),
\end{equation}
and similarly the survival probability for $\overline{\nu}_{\mu}$ is given by 
\begin{equation}
P(\overline{\nu}_{\mu} \rightarrow \overline{\nu}_{\mu}) \simeq 1-4\cos^{2}\overline{\theta}_{13}\sin^{2}\overline{\theta}_{23} \times [1-\cos^{2}\overline{\theta}_{13}\sin^{2}\overline{\theta}_{23}]\sin^{2}(\frac{1.267 |\Delta \overline{ m^{2}}_{32}|L}{E_{\overline{\nu}}}),
\end{equation}
where, the symbols with bar ($|\Delta\overline{ m^{2}}_{32}|, \overline{\theta}_{13}, \overline{\theta}_{23} $) are used for the respective parameters for the anti-neutrino oscillations.

In the Standard Model (SM) of particle physics, the parameters for particles and antiparticles are identical because of CPT symmetry. Hence, under the CPT symmetry, the mass splittings and mixing angles 
are identical for neutrinos and anti-neutrinos, implying $P(\nu_{\mu} \rightarrow \nu_{\mu})$ = $P(\overline{\nu}_{\mu} \rightarrow \overline{\nu}_{\mu})$. Any inequality between these disappearance probabilities of 
neutrinos and antineutrinos, could therefore, provide a hint for new physics. Also, any difference between the parameters governing the oscillation probabilities of neutrinos and antineutrinos could 
provide a possible hint for CPT violation. We investigate the prospects for the measurement of such a difference in $|\Delta m^{2}_{32}|$ and $|\Delta\overline{ m^{2}}_{32}|$ in Iron-Calorimeter (ICAL) experiment 
at the India-based Neutrino Observatory (INO) ~\cite{ino_tdr, INO3}. A similar work has been carried out by MINOS \cite{MINOS} and Super-Kamiokande \cite{SK} experiment. However, 
in the present work, for the first time, we show the INO-ICAL experimental sensitivity for the difference in $|\Delta m^{2}_{32}|$ and $|\Delta\overline{ m^{2}}_{32}|$ irrespective of the theoretical
mechanism responsible for the difference in the neutrino and antineutrino parameters.

The INO-ICAL experiment is an atmospheric neutrino experiment meant to study neutrino oscillations with muon disappearance channel. The ICAL experiment is sensitive to the atmospheric muon type neutrinos 
and antineutrinos through their interactions with the iron target producing muons and hadrons via the Charged Current (CC) interactions. The $\nu_\mu$ and $\overline{\nu}_\mu$ particles are identified by the 
production of $\mu^{-}$ and $\mu^{+}$, respectively, through their CC interaction ($\nu_\mu(\overline{\nu}_\mu) + X \rightarrow \mu^{-}(\mu^{+}) + X'$). The energy and direction of the incoming 
neutrinos/anti-neutrinos has to be measured accurately for the precise measurement of oscillation parameters. The energy and direction of these interacting neutrinos/anti-neutrinos can be determined 
from the reconstructed energy and direction of muons and hadrons. The muons deposit their energy in iron target forming a clear track-like pattern while hadrons form a shower-like pattern. 
Further details of the INO-ICAL experiment are provided in Sec.~\ref{ical_intro}. 

In this paper, we present the separate measurement of neutrino and anti-neutrino oscillation parameters. The INO-ICAL detector has a unique ability to distinguish between $\mu^{-}$ and $\mu^{+}$ events 
with their bendings in magnetic field, and hence can easily separate neutrinos and anti-neutrinos. Earlier ICAL analysis has shown its potential for measurement of oscillation parameters and mass 
hierarchy with combined neutrino and anti-neutrino events~\cite{anush, trk, DJ_epjc, moonmoon}. The reach of INO-ICAL experiment for CPT violation has also been studied in the litreature ~\cite{raj, poonam}. Here, we will analyze neutrino and anti-neutrino events separately and compare their oscillation parameters in order to find any signature of new physics including CPT violation~\cite{CPT1, CPT2}.
 
The analysis is performed with the construction of a $\chi^{2}$ function with 3D binning in muon energy, muon direction and hadron energy [Sec.~\ref{chianls}]. We calculate $\chi^{2}$ for neutrinos and 
antineutrinos separately and minimize it to constraint the experimental parameter spaces ($|\Delta m^{2}_{32}|$, $\theta_{23}$) and ($|\Delta\overline{m^{2}}_{32}|$, $\overline{\theta}_{23}$) [Sec.~\ref{identical_par}] assuming that the true parameters of 
$\nu_\mu$ and $\overline{\nu}_\mu$ are identical. Next, we study the possibility that the true values of $|\Delta m^{2}_{32}|$ and $|\Delta\overline{m^{2}}_{32}|$ can be different. In Sec.~\ref{diff_par}, 
we consider 4 theoretical scenarios when $|\Delta m^{2}_{32}|$ and $|\Delta\overline{ m^{2}}_{32}|$ take different set of values to find the $\Delta\chi^{2}$ contours for different experimental values of 
$|\Delta m^{2}_{32}|$ and $|\Delta\overline{ m^{2}}_{32}|$. This gives us the INO-ICAL sensitivity in the ($|\Delta m^{2}_{32}|$-$|\Delta\overline{m^{2}}_{32}|$) parameter space for these hypothetical 
scenarios.

The main aim of the paper is to find the sensitivity of INO-ICAL for the measurement of the difference ($|\Delta m^{2}_{32}|-|\Delta\overline{m^{2}}_{32}|$).
If the true values of the $|\Delta m^{2}_{32}|$ and $|\Delta\overline{m^{2}}_{32}|$ are different, at what confidence level the null hypothesis i.e. $|\Delta m^{2}_{32}|$=$|\Delta\overline{m^{2}}_{32}|$,
can be ruled out? This is addressed in Sec.~\ref{diff_sens}.

In this paper, we have demonstrated the potential of INO-ICAL for the separate measurement of $\nu_\mu$ and $\overline{\nu}_\mu$ oscillation parameters. 
The results of this study are summarized in Sec.~\ref{results}.

\section{The ICAL experiment}
\label{ical_intro}
India-based Neutrino Observatory (INO) is an approved mega science project, which will be located at Bodi West hills in the Theni district of South India. A huge
50 Kton ICAL detector with the magnetized iron target will operate at INO. The 1 km rock overburden above the site will act as a natural shield from the background of cosmic rays. 
The dimensions of the cavern will be $132m\times26m\times32m$. The ICAL detector will be of rectangular shape of dimensions $48m\times16m\times14.5m$ having three modules. Each module weighing about 
17 Kton with the dimensions $16m\times16m\times14.5m$. Each module will consist of 151 layers of 5.6~cm thick iron plates with alternate gaps of 4 cm where the active detector element will be placed. 
The ICAL experiment will use Resistive Plate Chambers as active detector element to detect the charged particles produced in the neutrino interaction with the iron nuclei. Since the INO experiment is 
expected to take data for several years to collect statistically significant number of interactions, the RPC’s are a good choice because of their long lifetime\cite{DJ_nim}. The RPC’s will give (X, Y) hit information with
0.96 cm spatial resolution. There will be a total of 30,000 RPC’s of dimension $2m\times2m$ in the ICAL detector. Another important feature of the INO-ICAL experiment is the application of a magnetic field 
of 1.5 T that will help in distinguishing the charge of the interacting particles. This distinction is crucial for the precise determination of relative ordering of neutrino mass states 
(neutrino mass hierarchy) and other parameters.

The INO-ICAL experiment is sensitive to atmospheric muons only. Hence, it will observe interactions of muon type neutrinos. The detector is not suitable for detection of electrons because of large thickness 
of iron plates (5.6~cm) compared to the radiation length of iron (17.6~cm). Also, tau lepton production is limited because of the high threshold of tau production (4~GeV). The ICAL experiment will also measure the energy of hadron shower to improve the energy reconstruction of events, and hence the overall sensitivity to neutrino parameters\cite{hdphy, DJ_epjc}.

The INO-ICAL resolutions for muon energy and direction as well as hadron energy are available from the GEANT4 \cite{geant} simulation studies~\cite{mureso, hreso}. The simulation studies provide us a reasonable characterization of the detector. 

\section{Analysis procedure}
The magnetized ICAL detector is primarily designed to differentiate the neutrino and anti-neutrino interactions of atmospheric muon neutrinos with excellent charge identification \cite{mureso}.
Since the events at ICAL can easily be separated into samples of $\nu_{\mu}$ and $\overline{\nu}_{\mu}$, they can be used to study oscillations separately in neutrinos and anti-neutrinos. We exploit this feature 
of ICAL experiment to measure the oscillation parameters independently using $\nu_{\mu}$ and $\overline{\nu}_{\mu}$  events assuming $|\Delta m^{2}_{32}|=|\Delta\overline{m^{2}}_{32}|$. Next, we explore the ICAL ability to 
find out any non-zero difference in the atmospheric mass squared differences of neutrinos and antineutrinos i.e. $|\Delta m^{2}_{32}|-|\Delta\overline{m^{2}}_{32}|$. 

For performing the analysis, we generated the atmospheric neutrino data set using HONDA et.al.\cite{honda} 3-D neutrino flux with ICAL detector 
specifications using NUANCE event generator~\cite{nuance}. For final $\chi^{2}$ analysis, 1000 years equivalent data of 50kt ICAL detector has been scaled down to 10 years exposure to normalize the  
statistical fluctuations. The ICAL detector is highly sensitive for the CC interactions of $\nu_{\mu}$ and $\overline{\nu}_{\mu}$ events in the energy range 0.8-12.8 GeV. Therefore, full event spectrum comprises 
the CC $\nu_{\mu}$ (and $\overline{\nu}_{\mu}$) events coming from $\nu_{\mu}(\overline{\nu}_{\mu}) \rightarrow \nu_{\mu}(\overline{\nu}_{\mu})$ survival channel and from 
$\nu_{e}(\overline{\nu}_{e}) \rightarrow \nu_{\mu}(\overline{\nu}_{\mu})$ oscillation channel. 
Initially, each event is generated without introducing oscillations, to reduce the computational time. The effect of oscillations have been incorporated separately using the Monte Carlo re-weighting algorithm 
described in earlier studies~\cite{trk, moonmoon,  DJ_epjc}. For each neutrino/antineutrino event of a given energy ($E_{\nu}$ or $E_{\overline{\nu}})$  and zenith direction $\theta_{z}$,  three flavor oscillation 
probabilities are calculated taking earth matter effects into account. The matter density profile of Earth is taken from the Preliminary Reference Earth Model \cite{prem} which divides the Earth into several
layers according to their matter densities.

In order to introduce the detector effects, we use the realistic detector resolutions and efficiencies of the ICAL detector based on GEANT4 simulations. The reconstruction of a neutrino 
(or anti-neutrino ) event requires the measurement of the secondary particles like muons (or anti-muon) and hadrons. The muons give clear track of hits inside the magnetized detector. Therefore, the energy 
of these particles can be easily reconstructed using a track fitting algorithm. The complete details of ICAL response for $\mu^{+}$ or $\mu^{-}$ e.g. energy and direction resolutions, reconstruction and 
charge identification efficiencies are available in Ref~\cite{mureso}. The ICAL has an excellent charge identification efficiency (more than 98$\%$) and good direction resolution for muons 
($\sim 1^{\circ}$) in the energy region of interest. Hadrons deposit their energies in a shower like pattern in the detector. So, total energy deposited by the hadron shower ($E^{\prime}_{had} = E_{\nu}-E_{\mu}$) 
is used to calibrate the detector response. The details of energy resolution and efficiency of hadrons at ICAL can be found in Ref \cite{hreso}. 

\subsection{The $\chi^{2}$ Function}
\label{chianls}

The oscillation parameters of the atmospheric neutrinos have been extracted with a $\chi^{2}$ analysis. The re-weighted events, with detector resolutions and efficiencies folded in, are binned into
the observed muon energy, muon direction and hadron energy. An optimized bin width have been used for these observables to get 
statistically significant event rates. The data has been divided into a total of 20 muon energy bins and 5 hadron energy bins with varying bin widths. A total of 20 muon direction bins for 
$\cos\theta_{\mu}$ in the range of -1 to 1, with equal bin width has been chosen. 
The above mentioned binning scheme is applied for both $\nu_{\mu}$ and $\overline{\nu}_{\mu}$ events. The details of the binning scheme is shown in Table~\ref{bin_tb}. 
 \begin{table}
\begin{center}
\begin {tabular}{c c c}
\hline
\hline
 Muon energy bins ($E_{\mu^{\pm}}$ in GeV) & Range & Bin width  \\
\hline 
12 &  0.8-4.8 & 0.34 \\
 4 &  4.8-6.8 &  0.5\\
3 &  6.8-9.8& 1 \\
1&  9.8-12.8 & 3\\
 \hline
 \hline

Hadron energy bins ($E_{hadron}$ in GeV) &  &   \\
 \hline
2 &  0.0-2.0 & 1 \\
2&  2.0-8.0 & 3 \\
1&  8.0-13 & 5\\
\hline
 \hline

Muon angle bins ($\cos\theta_{\mu^{\pm}}$) &  & \\
\hline 
20 & -1 - +1  & 0.1  \\
 \hline
\end {tabular}
\caption{\label{bin_tb} Optimized 3D binning scheme used for analyses.}                                      
\end{center}
\end{table}

A ``pulled'' $\chi^{2}$ \cite{maltoni} method based on Poisson probability distribution is used to compare the expected and observed data with inclusion of systematic errors. Five systematic errors 
used in analysis are: a 20\% error on atmospheric neutrino flux normalization, 10\% error on neutrino cross-section, an overall 5\% statistical error, a 5\% uncertainty due to zenith angle dependence of 
the fluxes, and an energy dependent tilt error, as considered in earlier ICAL analyses \cite{trk, DJ_epjc}.  

 In the method of pulls, systematic uncertainties and the theoretical errors are parameterized in terms of a set of variables $\zeta$, called pulls.  Due to the fine binning, some bins may have very 
 small number of entries. Therefore, we use the poissonian definition of $\chi^{2}$ given as
 
 \begin{equation}
\label{eq:chieq}
 \chi^2(\nu_{\mu}) = min\sum_{i,j,k}\left(2 (N^{th^{\prime}}_{ijk}(\nu_{\mu})-N^{ex}_{i,j,k}(\nu_{\mu}))+2N^{ex}_{i,j,k}(\nu_{\mu})
(\ln\frac{N^{ex}_{i,j,k}(\nu_{\mu})}{N^{th^{\prime}}_{i,j,k}(\nu_{\mu})})\right)+ \sum_{n}\zeta^{2}_{n}, 
 \end{equation}
 where 
\begin{equation}
\label{eq:evteq}
  N^{th^{\prime}}_{ijk}(\nu_{\mu}) = N^{th}_{i,j,k}(\nu_{\mu})\left(1 + \sum_{n}\pi^{n}_{ijk}\zeta_{n}\right). 
 \end{equation}
 Here, $N^{ex}_{ijk}$ are the observed number of reconstructed events, generated using true values of the oscillation parameters  in $i^{th}$ muon energy bin, $j^{th}$ muon direction bin and $k^{th}$ hadron 
 energy bin.  In Eq.(~\ref{eq:evteq}), $N^{th}_{ijk}$ are the number of theoretically predicted events generated by varying oscillation parameters, $N^{th^{\prime}}_{ijk}$ show modified events spectrum due to
 different systematic uncertainties, $\pi^{n}_{ijk}$ are the systematic shift in the events of the respective bins due to $n^{th}$ systematic error. The variable $\zeta_{n}$, the univariate pull variable,
 corresponds to the $\pi^{n}_{ijk}$ uncertainty. 
An expression similar to Eq.(~\ref{eq:chieq}) can be obtained for $\chi^{2}(\overline{\nu}_{\mu})$ using reconstructed $\mu^{+}$ event samples.  

The functions 
$\chi^2(\nu_{\mu})$ and $\chi^2(\overline{\nu}_{\mu})$ are calculated separately for the independent measurement of neutrino and anti-neutrino oscillation parameters. The two $\chi^{2}$ can be added to get the 
combined
$\chi^2(\nu_{\mu}+\overline{\nu}_{\mu})$ as
 \begin{equation}
\label{eq:chiino}
  \chi^2(\nu_{\mu}+\overline{\nu}_{\mu}) =\chi^{2}(\nu_{\mu}) + \chi^{2}(\overline{\nu}_{\mu}).
 \end{equation}
 
\subsection{Same True oscillation parameters for neutrinos and antineutrinos}
 \label{identical_par}
In the present work, we investigate the scenario where the neutrino and antineutrino oscillations are different. However, we begin with the case where neutrinos and antineutrinos have identical 
oscillation parameters ($|\Delta m^{2}_{32}|= |\Delta\overline{m^{2}}_{32}|$, $\sin^{2}\theta_{23} =\sin^{2}\overline{\theta}_{23}$). The central true values of the oscillation parameters and their marginalization range used in the analysis are shown in Table \ref{osc_tb}. The $\chi^{2}$ have been calculated as the function of the atmospheric oscillation parameters ( $|\Delta m^{2}_{32}|$ and $ \sin^2\theta_{23}$) while all other 
oscillation parameters are kept fixed at their central values.
 The  solar oscillation parameters $\Delta m^{2}_{21}$ and $ \sin^{2}\theta_{12}$ are fixed as they do not show significant impact on the results. Since $\theta_{13}$ is now known quite precisely, 
 it has been kept fixed as well. Since, ICAL is not sensitive to the $\delta_{CP}$ \cite{cpsense}, it is kept fixed at $0^{\degree}$. 
 \begin{table}
\begin{center}
\begin {tabular}{c  c  c}
\hline
\hline
Oscillation parameters & True values & Marginalization range  \\
\hline 
\hline
$ \sin^2(2\theta_{12})$ &  0.86 & Fixed \\
$ \sin^2(\theta_{23})$ &  0.5 &  0.4-0.6 \\
$ \sin^2(\theta_{13})$ &  0.0234 &  Fixed                 \\
$\Delta m^{2}_{21}$ (eV$^{2}$) &  7.6 $\times$ $10^{-5}$& Fixed\\
 $|\Delta m^{2}_{32}|$ (eV$^{2}$) & 2.4 $\times$ $10^{-3}$ & (2.1-2.6) $\times$ $10^{-3}$\\
 $\delta$ & 0.0 & Fixed \\
 \hline
\end {tabular}
\caption{\label{osc_tb} True values of the neutrino oscillation parameters used in the analysis. We vary $\sin^{2}\theta_{23}$ and $|\Delta m^{2}_{32}|$ in their $3\sigma$ range whereas the other 
vairables are kept fixed.} 
\end{center}
\end{table}

 In order to obtain the experimental sensitivity for $\sin^{2}\theta_{23}$ and $|\Delta m^{2}_{32}|$, we independently minimize the   $\chi^2(\nu_{\mu})$, $\chi^2(\overline{\nu}_{\mu})$ and combined  
 $\chi^2(\nu_{\mu}+\overline{\nu}_{\mu})$ function by varying oscillation parameters within their allowed ranges with all systematic uncertainties folded in. The precision on the oscillation parameters can be defined 
 as the ratio of $(P_{max}-P_{min})$ to the $(P_{max}+P_{min})$, where $P_{max}$ and $P_{min}$ are the maximum and minimum values of the concerned oscillation parameters at the given confidence level.

\begin{figure}[ht]
\centering
  \includegraphics[width=0.5\textwidth,height=6cm]{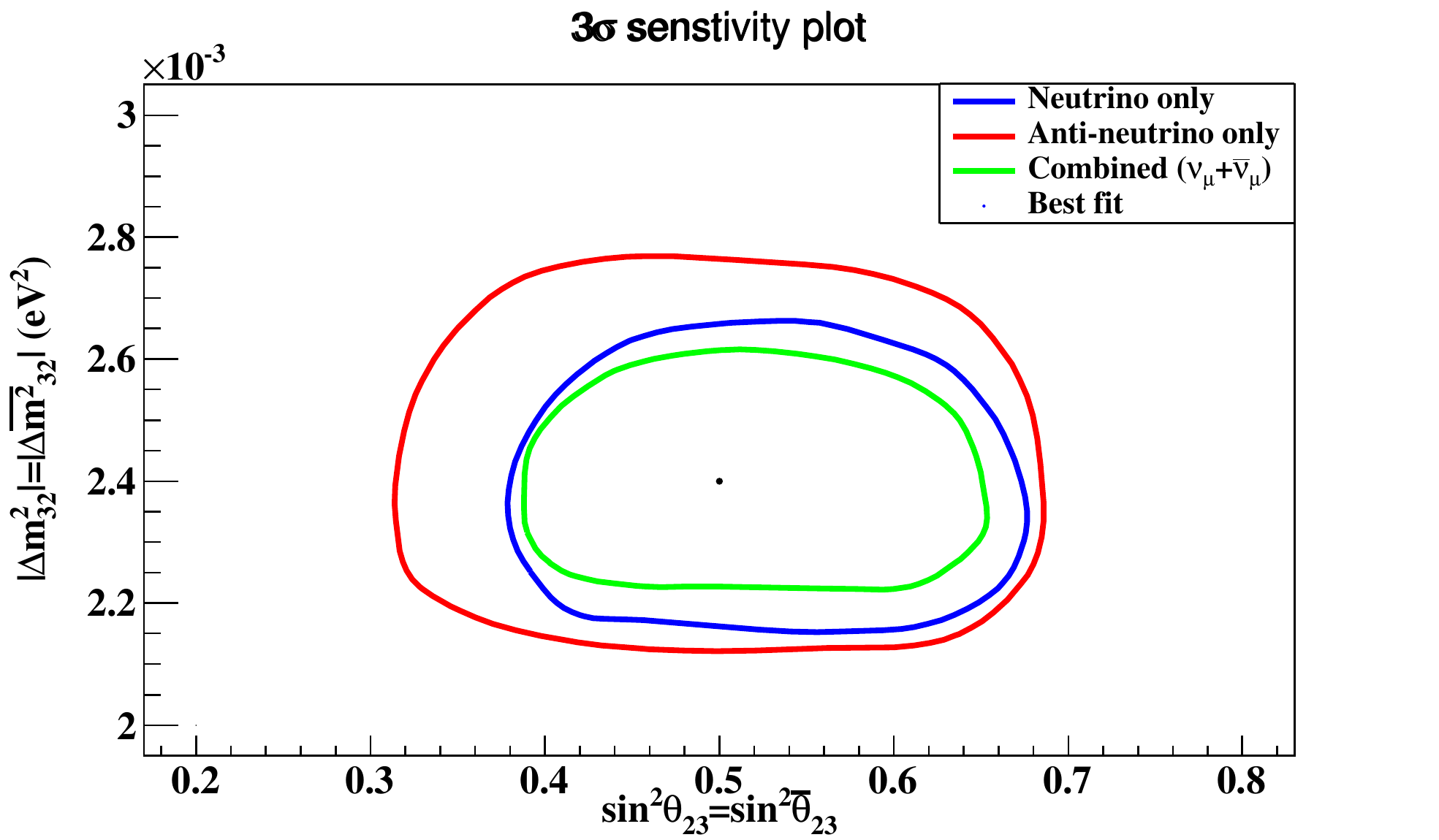}
  \caption{The 3$\sigma$ sensitivity plot on ($\sin^{2}\theta_{23}$=$\sin^{2}\overline{\theta}_{23}$, $|\Delta m^{2}_{32}|$=$|\Delta \overline{m^{2}}_{32}|$) parameter space from the $\chi^2$ analyses
  separately for neutrino only events, antineutrino only events and combined neutrino and antineutrino ($\nu_{\mu}+\overline{\nu}_{\mu}$) events}\label{fig:3sig}
\end{figure}

 Fig.\ref {fig:3sig}  shows the resulting contours at $3\sigma$ confidence level (C.L.) obtained for the $(|\Delta m^{2}_{32}|$, $\sin^{2}\theta_{23})$ or $(|\Delta\overline{m^{2}}_{32}|, \sin^{2}\overline{\theta}_{23}) $ planes. 
 These results are also compared with combined results of neutrino and antineutrino events.
 \begin{table}
\begin{center}
\begin {tabular}{c c c}
\hline
\hline
Analysis & $ \sin^2\theta_{23}$ &$|\Delta m^{2}_{32}|$ (eV$^{2}$)  \\
\hline 
\hline
Neutrino events &  27.1$\%$ &10.4$\%$ \\
Anti-neutrino events &  38.0$\%$ & 13.4$\%$\\
Combined($\nu_{\mu}+\overline{\nu}_{\mu}$)&  25.0$\%$ & 8.3$\%$\\
 \hline
\end {tabular}
\caption{\label{pre_tb}Precision values at the $3\sigma$ C.L. considering $\sin^2\theta_{23}=\sin^2\overline{\theta}_{23}$ and $|\Delta m^{2}_{32}|$=$|\Delta\overline{m^{2}}_{32}|$ (eV$^{2}$) obtained from neutrino only events, anti-neutrino only events and with the combined ($\nu_{\mu} +\overline{\nu}_{\mu}$) events. } 
\end{center}
\end{table}

 Table~\ref{pre_tb} shows the precision values at the $3\sigma$ C.L. obtained from neutrino only events, anti-neutrino only events and with the combined ($\nu_{\mu} +\overline{\nu}_{\mu}$) events.
It can be observed that combined $\nu_{\mu}$ and $\overline{\nu}_{\mu}$ analysis gives more precise values of the oscillation parameters, as expected. The major contribution to this precision
 comes from the higher statistics of the combined neutrino events. However, the ICAL detector can also  measure $|\Delta m^{2}_{32}|$ very precisely from  neutrino and antineutrino events, separately.
 It can be noted that the 
 allowed parameter space of anti-neutrino analysis is wider than the neutrino only events analysis due to their low statistics.  

\subsection{Different True values of $|\Delta m^{2}_{32}|$ and $|\Delta\overline{m^{2}}_{32}|$ }
\label{diff_par}
The good precision of INO-ICAL for $|\Delta m^{2}_{32}|$ and $|\Delta\overline{m^{2}}_{32}|$ motivates us to examine the scenario when the true values of $|\Delta m^{2}_{32}|$ and $|\Delta\overline{m^{2}}_{32}|$
have different values. This will allow us to either establish or rule out the hypothesis that neutrinos and antineutrinos have same oscillation parameters.

\begin{figure}[htbp]
 \centering
 \subfigure[]{
  \includegraphics[width=0.45\textwidth,height=6cm]{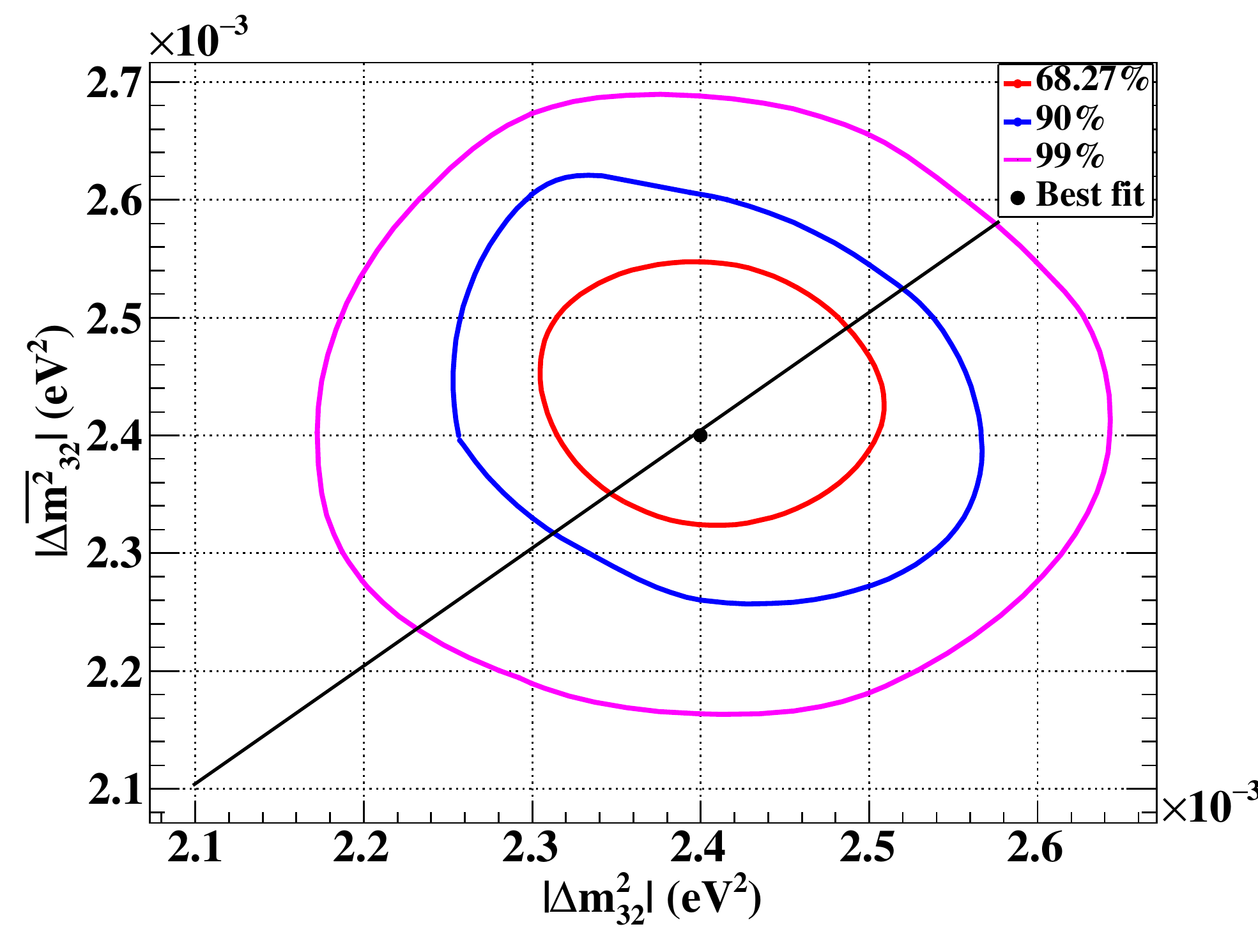}
   \label{fig:a}
   }
 \subfigure[]{
  \includegraphics[width=0.45\textwidth,height=6cm]{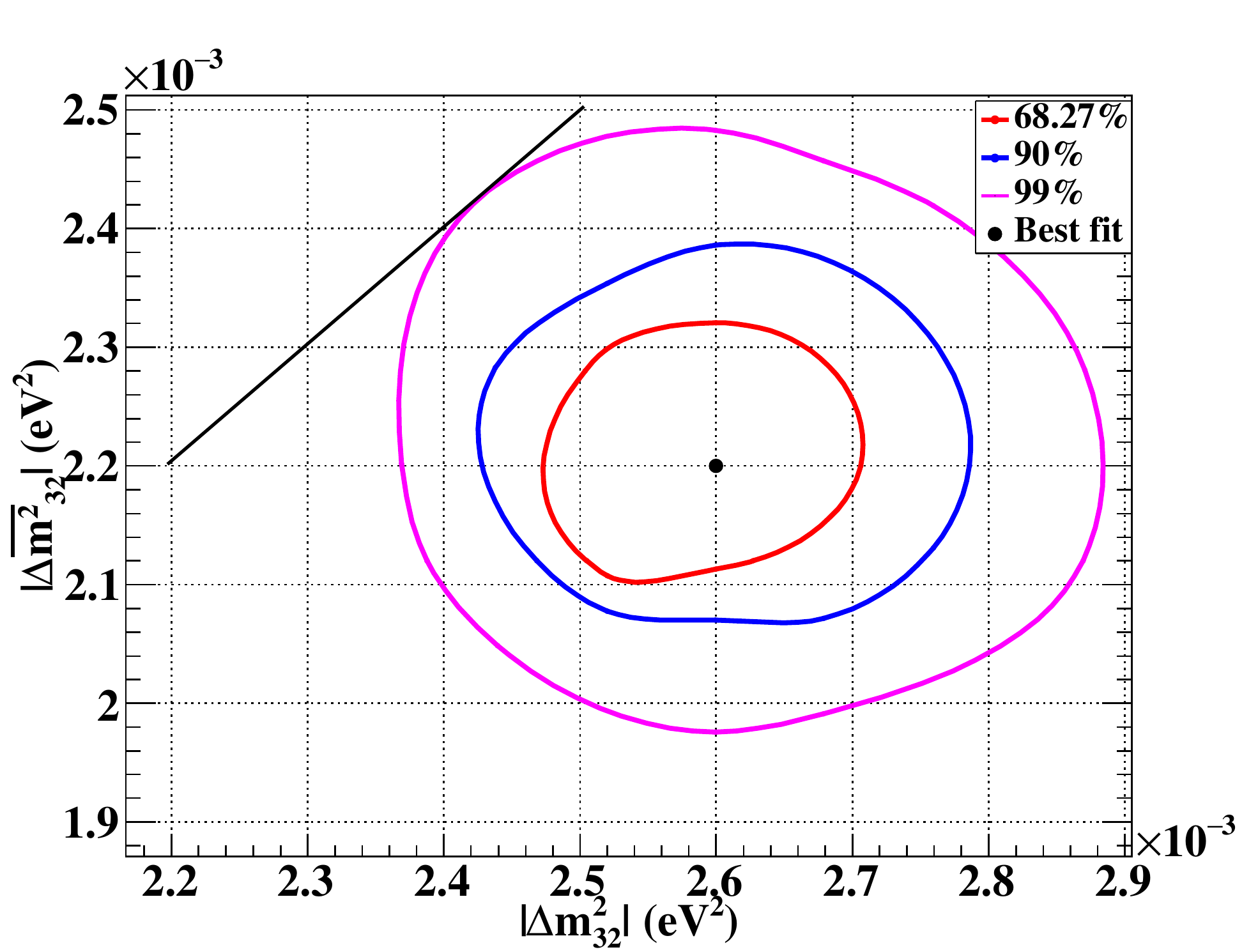}
   \label{fig:b}
   }
\subfigure[]{
  \includegraphics[width=0.45\textwidth,height=6cm]{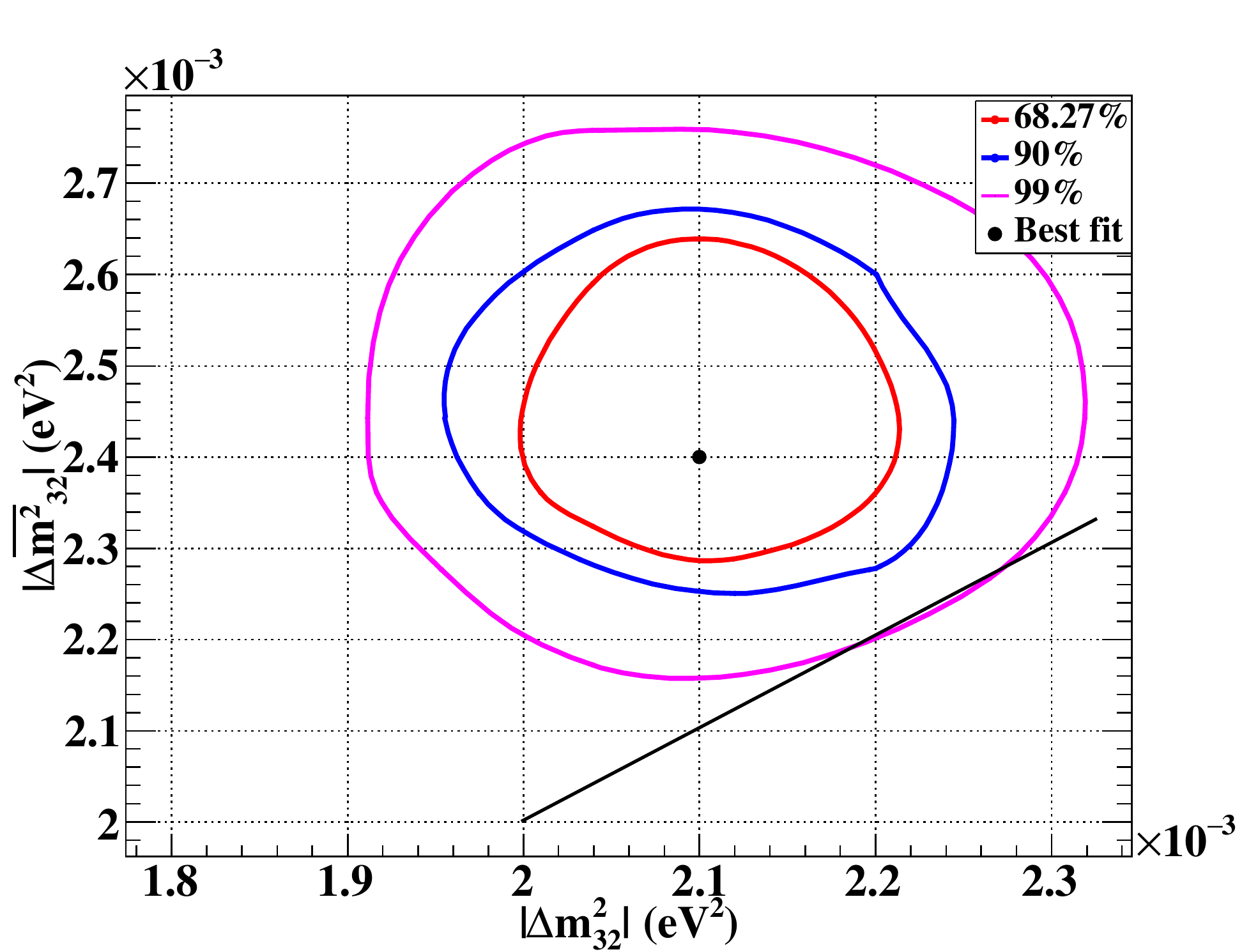}
   \label{fig:c}
   }
\subfigure[]{
  \includegraphics[width=0.45\textwidth,height=6cm]{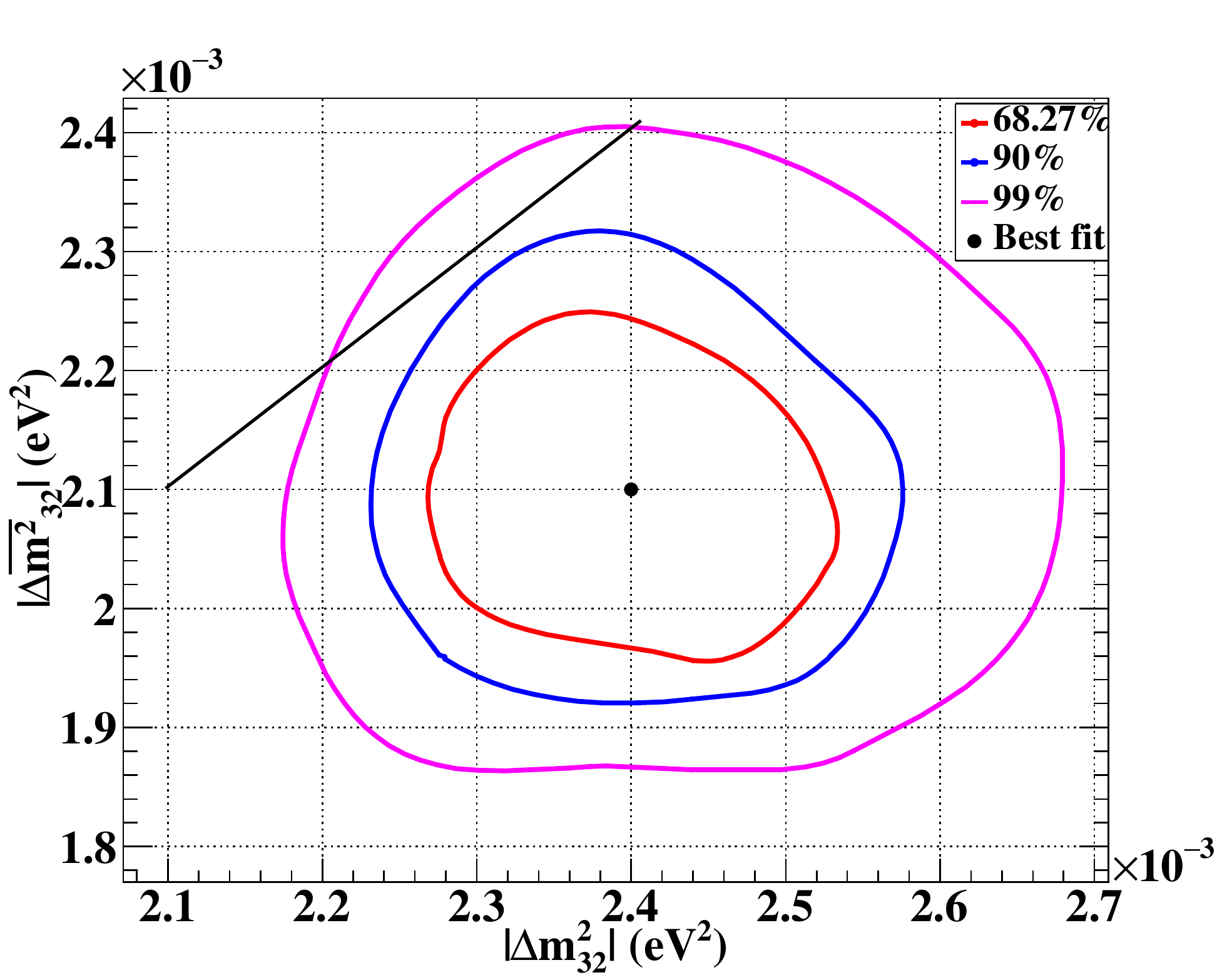}
   \label{fig:d}
   }
\caption{\label{sample}Contour plots at 68$\%$, 90$\%$ and 99$\%$ C.L. for different true values of $|\Delta m^{2}_{32}|$ and $ |\Delta\overline{m^{2}}_{32}|$ as mentioned in Table ~\ref{par_tb}. 
Here, X-axis corresponds to $|\Delta m^{2}_{32}|$ values and Y-axis corresponds to $|\Delta\overline{m^{2}}_{32}|$ values.}
\end{figure}

We assume that neutrinos and antineutrinos have different true values of mass squared differences ($|\Delta m^{2}_{32}|$, $|\Delta\overline{m^{2}}_{32}|$).
All other oscillation parameters are same as in the previous section. We take different representative cases of the true values of $|\Delta m^{2}_{32}|$ and $|\Delta\overline{m^{2}}_{32}|$ 
and estimate $\chi^2$ as a function of the $|\Delta m^{2}_{32}|$ and $|\Delta\overline{m^{2}}_{32}|$. For each case, the true values of all oscillation parameters are fixed and
$\chi^2(\nu+\overline{\nu})$ have been estimated as a function of observed values of $|\Delta m^{2}_{32}|$ and $\Delta\overline{m^{2}}_{32}|$. The $\chi^2$ contours at 68$\%$, 90$\%$ and 99$\%$ C.L. have been 
plotted on the ($|\Delta m^{2}_{32}|$, $|\Delta\overline{m^{2}}_{32}|$) parameter space. The straight line corresponding to the null hypothesis 
($|\Delta\overline{m^{2}}_{32}|$=$|\Delta m^{2}_{32}|$) is also shown. If the null hypothesis line is n$\sigma$ away form the $\chi^{2}$ minimum, it can be concluded that
the null hypothesis ($|\Delta\overline{m^{2}}_{32}|$=$|\Delta m^{2}_{32}|$) is ruled out at n$\sigma$ C.L. The four plots in Fig.~\ref{sample} correspond to the true values 
of  $|\Delta m^{2}_{32}|$ and  $|\Delta\overline{m^{2}}_{32}|$ as shown in Table~\ref{par_tb}. 

 \begin{table}[htbp]
\begin{center}
\begin {tabular}{c c c c}
\hline
\hline
\\[-1em]
 Fig2.No.&$|\Delta m^{2}_{32}|$ (eV$^{2}$) & $|\Delta\overline{m^{2}}_{32}|$ (eV$^{2}$) &  $|\Delta m^{2}_{32}|$-$|\Delta\overline{m^{2}}_{32}|$ (eV$^{2}$)\\
\hline
 \hline
(a)&2.4$\times 10^{-3}$ & 2.4$\times 10^{-3}$ & 0.0$\times 10^{-3}$ \\
(b)&2.6$\times 10^{-3}$ & 2.2$\times 10^{-3}$ & 0.4$\times 10^{-3}$\\
(c)&2.1$\times 10^{-3}$&  2.4$\times 10^{-3}$ & -0.3$\times 10^{-3}$\\
(d)&2.4$\times 10^{-3}$&  2.1$\times 10^{-3}$ & +0.3$\times 10^{-3}$\\
 \hline
\end {tabular}
\caption{\label{par_tb} Different combinations of $|\Delta m^{2}_{32}|$ and $|\Delta\overline{m^{2}}_{32}|$ values used in Fig~\ref{sample}.} 
\end{center}
\end{table}

Fig.\ref{fig:a} show the contours when true values of $|\Delta m^{2}_{32}|$ and $ |\Delta\overline{m^{2}}_{32}|$ are exactly equal 
(i.e.$|\Delta m^{2}_{32}|-|\Delta\overline{m^{2}}_{32}|=0$). In this case, the null hypothesis line ( solid black line) crosses the central best fit point. In Fig.~\ref{fig:b}, \ref{fig:c} and \ref{fig:d} the difference 
(i.e.$|\Delta m^{2}_{32}|-|\Delta\overline{m^{2}}_{32}|$) is non-zero, and hence the best fit point shifts away from the null hypothesis line.
Fig.\ref{fig:b} correspond to the case when true values of $|\Delta m^{2}_{32}| =2.6\times 10^{-3} eV^{2}$ and $ |\Delta\overline{m^{2}}_{32}|= 2.2\times 10^{-3} eV^{2}$ 
(i.e.$|\Delta m^{2}_{32}|-|\Delta\overline{m^{2}}_{32}|=  0.4\times 10^{-3} eV^{2}$). Here, the null hypothesis line is tangential to  $3\sigma$ contour. So, the tangential point 
is $3\sigma$ away from the central best fit value. Thus, it can be concluded that null hypothesis is ruled out at $3\sigma$ C.L. Similarly, Fig.~\ref{fig:c} shows that null hypothesis is tangential to $3\sigma$ CL 
when true values of $|\Delta m^{2}_{32}| =2.1\times 10^{-3} eV^{2}$ and $ |\Delta\overline{m^{2}}_{32}|= 2.4\times 10^{-3} eV^{2}$ 
(i.e.$|\Delta m^{2}_{32}|-|\Delta\overline{m^{2}}_{32}|= -0.3\times 10^{-3} eV^{2}$). Fig.\ref{fig:d} shows that the null hypothesis is ruled out at roughly $2.5\sigma$ CL 
when true value of $|\Delta m^{2}_{32}| =2.4\times 10^{-3} eV^{2}$ and $ |\Delta\overline{m^{2}}_{32}|= 2.1\times 10^{-3} eV^{2}$ (i.e.$|\Delta m^{2}_{32}|-|\Delta\overline{m^{2}}_{32}|= +0.3\times 10^{-3} eV^{2}$).

\subsection {ICAL sensitivity for $|\Delta m^{2}_{32}|$--$|\Delta\overline{m^{2}}_{32}|\neq 0$}
\label{diff_sens}
\begin{figure}[htbp]
 \centering
   \includegraphics[width=1.0\textwidth,height=12cm]{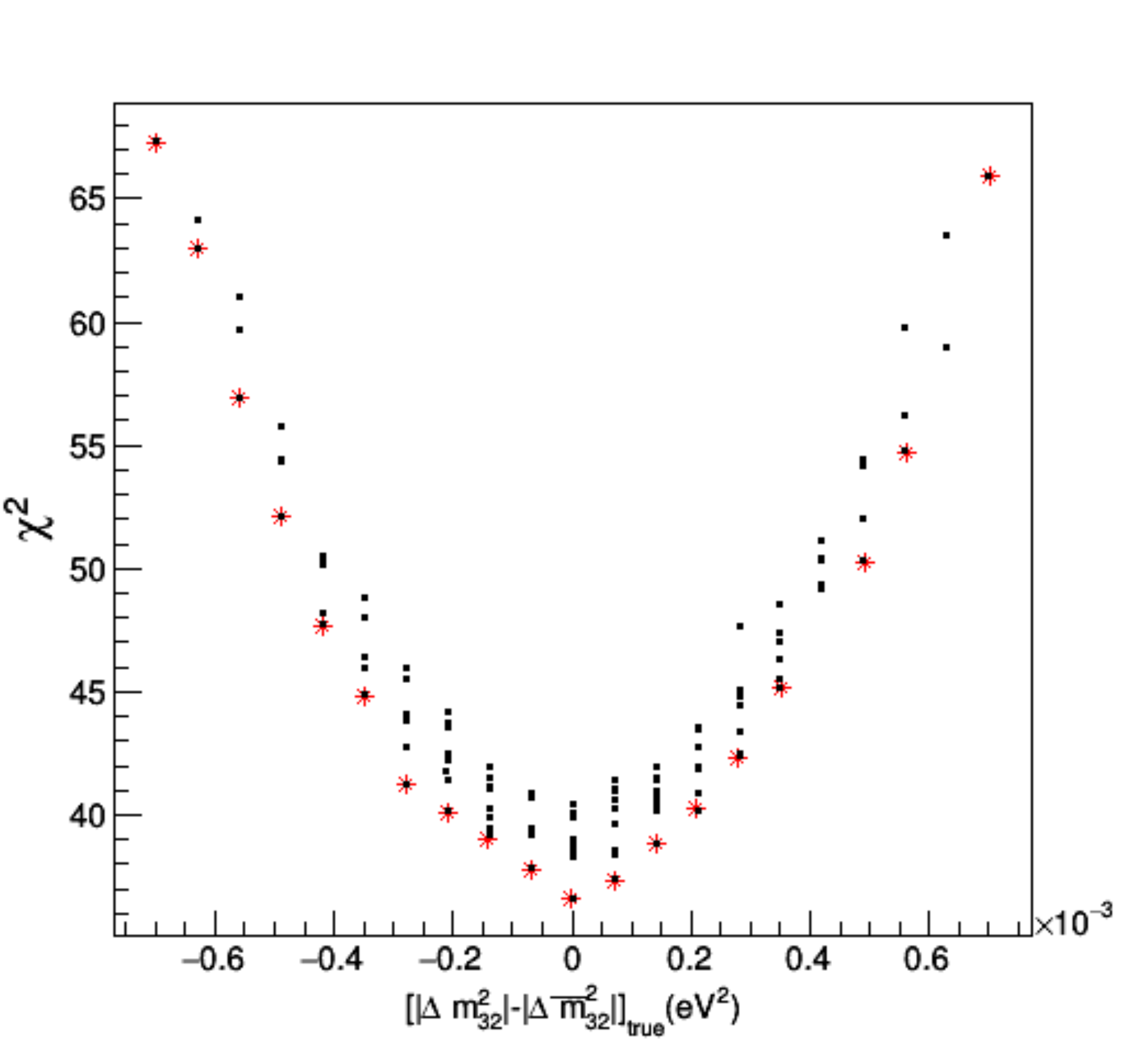}
   \caption{$\chi^{2}$ versus $(|\Delta m^{2}_{32}|-|\Delta\overline{m^{2}}_{32}|)_{True}(eV^{2})$ plot, black dots represents the several minimum $\chi^2$ for a 
   common ($|\Delta m^{2}_{32}|-|\Delta\overline{m^{2}}_{32}|$) difference and red stars depicts the  smallest $\chi^2$ value among all of them.}\label{trueplot}
 \end{figure}

In order to check the ICAL sensitivity for a non-zero value of the difference between $|\Delta m^{2}_{32}|$ and  $|\Delta\overline{m^{2}}_{32}|$, the true values of $|\Delta m^{2}_{32}|$ and  
$|\Delta\overline{m^{2}}_{32}|$ have been varied independently in a range ($0.0021-0.0028 eV^{2}$). But, we estimate the $\chi^{2}(\nu+\overline{\nu})$ only when the observed values of $|\Delta m^{2}_{32}|$
and  $|\Delta\overline{m^{2}}_{32}|$ are equal. In other words, the $\chi^{2}(\nu+\overline{\nu})$ is being estimated on the null hypothesis
line where the $|\Delta m^{2}_{32}|$ and  $|\Delta\overline{m^{2}}_{32}|$ values are equal. The minimum value of $\chi^2$ is chosen on this line that corresponds to the tangential point
where the null hypothesis line coincides with the corresponding contour. Finally, this minimum $\chi^2$ is binned as a function of difference in the true values of 
($|\Delta m^{2}_{32}|$-$|\Delta\overline{m^{2}}_{32}|$). This  will result in several $\chi^2$ points corresponding to a common ($|\Delta m^{2}_{32}|-|\Delta\overline{m^{2}}_{32}|$) difference,
depicted as dots in Fig.\ref{trueplot}. From all such candidate points, we pick those points that have the smallest $\chi^2$ values and depict them as stars [see Fig.\ref{trueplot}].

\begin{figure}[htbp]
 \centering
   \includegraphics[width=1.0\textwidth,height=12cm]{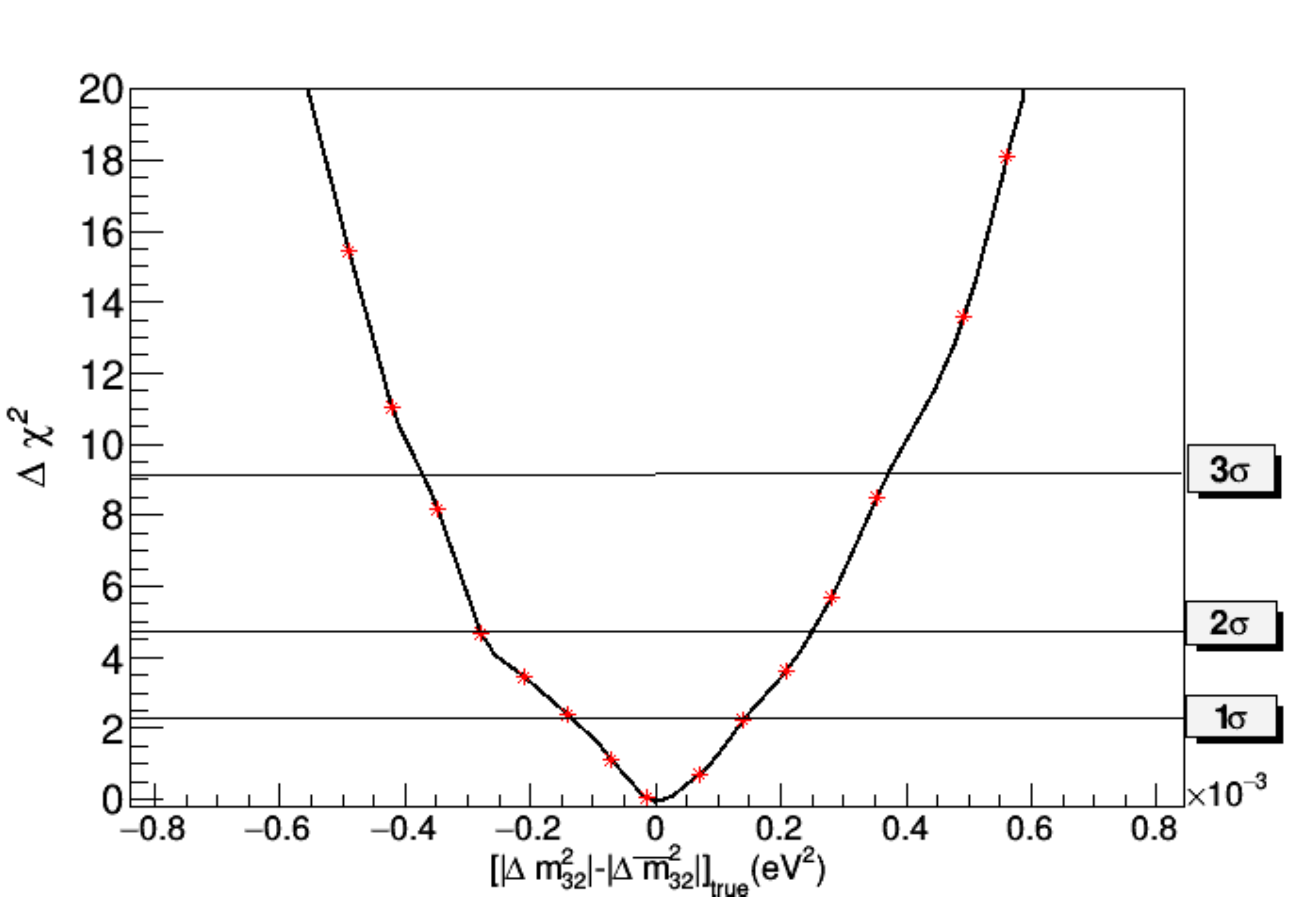}
   \caption{The INO-ICAL sensitivity for $(|\Delta m^{2}_{32}|-|\Delta\overline{m^{2}}_{32}|)_{True}(eV^{2})$ at 1$\sigma$, 2$\sigma$ and 3$\sigma$ confidence levels.}
\label{trueplot2}
\end{figure}

For each value of  ($|\Delta m^{2}_{32}|-|\Delta\overline{m^{2}}_{32}|$), we calculate $\Delta\chi^{2}=\chi^{2}-\chi^{2}_{min}$ assuming $\chi^2_{min}=37$ and plot it as the functions of 
$|\Delta m^{2}_{32}|-|\Delta\overline{m^{2}}_{32}|$ in Fig~\ref{trueplot2}.
This figure depicts the INO-ICAL potential for ruling out the null hypothesis $|\Delta m^{2}_{32}|=|\Delta\overline{m^{2}}_{32}|$ and is our final result of the present study.

\section {Results and Conclusions}
\label{results}
The experimental confirmation of different sets of oscillation parameters for neutrinos and antineutrinos will be a signature of any new physics like CPT symmetry in the neutrino sector. In this paper, we investigate 
this possibility by studying INO-ICAL potential for the separate measurements of neutrino and antineutrino oscillation parameters for 10 years of exposure. The CC $\nu_{\mu}$ and $\overline{\nu}_{\mu}$ events 
are separated into muon energy, muon direction and hadron energy bins. A $\chi^{2}$ analysis is used with
realistic detector resolutions, efficiencies and systematic errors. The separate analysis for neutrino and antineutrino events having identical oscillation parameters 
($|\Delta m^{2}_{32}|=|\Delta\overline{m^{2}}_{32}|, \sin^{2}\theta_{23}= \sin^{2}{\overline{\theta}_{23}}$) indicates that ICAL can measure the atmospheric neutrino 
parameters $|\Delta m^{2}_{32}|$ with a precision of 10.14$\%$ and $\sin^{2}\theta_{23}$ with a precision of 27.10$\%$. The atmospheric antineutrino 
parameters $|\Delta\overline{m^{2}}_{32}|$ and $\sin^{2}\overline{\theta}_{23}$ can be measured with a precision of 13.4$\%$ and 38.0$\%$ at 3$\sigma$ confidence level respectively. 
As expected, the combined $\nu_{\mu}$+$\overline{\nu}_{\mu}$ events show a better sensitivity with a precision of 8.7$\%$ for 
$|\Delta m^{2}_{32}|$ and 25.0$\%$ for $\sin^{2}\theta_{23}$ at same confidence level due to larger events in $\chi^{2}$.

Further, we investigate the scenario where the neutrino and antineutrino oscillation parameters have different values. We measure the ICAL sensitivity for ruling out the null hypothesis 
$(|\Delta m^{2}_{32}|=|\Delta\overline{m^{2}}_{32}|)$ by estimating the difference between the true values of mass squared differences of neutrinos and antineutrinos i.e. 
$(|\Delta m^{2}_{32}|-|\Delta\overline{m^{2}}_{32}|)$. We show that ICAL can rule out the null hypothesis of $|\Delta m^{2}_{32}|=|\Delta\overline{m^{2}}_{32}|$ at more than 
3$\sigma$ level if the difference of true values of $|\Delta m^{2}_{32}|-|\Delta\overline{m^{2}}_{32}|\geq +0.4\times10^{-3}eV^{2}$ or $|\Delta m^{2}_{32}|-|\Delta\overline{m^{2}}_{32}|\leq -0.4\times10^{-3}eV^{2}$.


\section{Acknowledgment}
The authors would like to thank Department of Science and Technology (DST), Govt. of India for their generous funding support. We would also like to thank University of Delhi for providing R $\&$ D grants 
to carry out a part of this work.

\end{document}